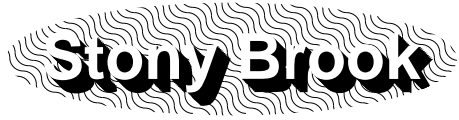

# SELF-DUAL N=8 SUPERGRAVITY
# AS CLOSED N=2(4) STRINGS

W. Siegel[1]

*Institute for Theoretical Physics*
*State University of New York, Stony Brook, NY 11794-3840*

**ABSTRACT**

As open N=2 or 4 strings describe self-dual N=4 super Yang-Mills in 2+2 dimensions, the corresponding closed (heterotic) strings describe self-dual ungauged (gauged) N=8 supergravity. These theories are conveniently formulated in a chiral superspace with general supercoordinate and local OSp(8|2) gauge invariances. The super-light-cone and covariant-component actions are analyzed. Because only half the Lorentz group is gauged, the gravity field equation is just the vanishing of the torsion.

[1] Work supported by National Science Foundation grant PHY 90-08936.
Internet address: siegel@dirac.physics.sunysb.edu.

# 1. Results

N=2 strings [1] have been proposed as theories of self-dual Yang-Mills and gravity [2-4]. In a previous paper [5] we showed how the open string [4] actually describes self-dual *super* Yang-Mills, with all the helicities from $+1$ to $-1$. The amplitudes of this field theory in an appropriate gauge are identical to those found in the string theory, but Lorentz invariance shows that the external states have helicities other than just the $+1$ of nonsupersymmetric self-dual gluons. Previously the states of other helicities had been neglected because spectral flow indicated the equivalence of states with different boundary conditions [6]. This is obviously not the case if some of the states are fermions. The fermionic contributions to all loop corrections cancel those of the bosons because of the trivial nature of the supersymmetry generated by spectral flow [5]. (This avoids some apparent problems from infrared divergences [7] when the fermions are neglected.) This residual symmetry of the N=2 string U(1) constraint is analogous to the residual global U(1) symmetry which exists in QED in spite of the local U(1) constraint (the quantum version of Coulomb's law) which appears in quantization in the temporal gauge (the analog of the string's conformal gauge). The fact that some of the states are fermions is obscured in the N=2 formalism for this string because the SO(2,2) Lorentz symmetry is not manifest. (In fact, it was not realized until ref. [2] that the theory contained Yang-Mills and gravity and not just scalars.) However, in the equivalent (world-sheet) N=4 formulation, where this Lorentz symmetry is manifest, there are clearly spinors [8]. In [5] we attacked this problem from the field theory approach to avoid the complications of quantization of the N=4 string. In this paper we continue this approach for the closed [2] and heterotic [3] strings, which describe self-dual ungauged and gauged N=8 supergravity.

From the field theory point of view, the appearance of states other than gluons and/or gravitons follows from the requirement of a Lorentz covariant action. The self-duality conditions are then enforced by Lagrange multipliers. The existence of these Lagrange multipliers is implied by supersymmetry: They are in the same supersymmetric multiplet as Yang-Mills/gravity, but only when the supersymmetry is maximal (N=4 for super Yang-Mills, N=8 for supergravity), since Lorentz invariance requires the multiplier have helicity equal in magnitude but opposite in sign to the physical self-dual Yang-Mills/gravity polarization. (This is also why for SO(2,2) all helicities, except sometimes helicity zero, come with indefinite Hilbert space metric: Opposite helicities are not complex conjugates, but are both real.) This is particularly clear in the light-cone formalism, where only propagating degrees of freedom appear, and the field content follows directly from the unique free action $\int d^4x\ d^N\theta\ \frac{1}{2}V\Box V$. The fact that all states are in the same supersymmetric multiplet is the statement that all states (boundary conditions) of the string are related by spectral flow. The minimal off-shell field content, and the Lorentz and gauge covariant form of the interactions, is determined from the commutation relations of the covariant derivatives, which can also be solved explicitly in the light-cone gauge. The light-cone form of the interaction vertices for open and closed strings is independent of the helicities of the states, another consequence of spectral flow, which is seen here to follow from just supersymmetry and self-duality.

Thus, Lorentz covariance, supersymmetry, and self-duality uniquely determine the open string to be self-dual N=4 super Yang-Mills and the closed string to be self-dual N=8 supergravity. The different values of N prevent the closed and open strings from coupling, but this is also implied from the vanishing of the usual 1-loop string diagrams which generate closed strings from open ones. (This differs from the result of [4], where open and closed strings are coupled, since in the absence of fermions 1-loop diagrams survive, as well as infrared divergences.) However, the heterotic string is more complicated, since even when spectral flow is taken into account there are still distinct gluons and gravitons both in the theory, as well as the two corresponding independent coupling constants. (Although the heterotic string apparently requires compactification to 2 or 3 dimensions [3], here we consider only the 2+2-dimensional theory that gives that theory upon dimensional reduction.) The solution also follows from supersymmetry: Gauged N=8 supergravity has two couplings, including one for the nonabelian gauge vectors. (The N=8 supersymmetry is again required by the appearance of the graviton.) This result is again unique, up to the choice of gauge group for the 28 vectors (SO(n,8-n) or one of its contractions). This is smaller than the number of vectors expected from the string analysis [3], but there may be some analog of GSO projection at work: In particular, this is the spectrum expected for the N=(2,1) heterotic string for such a projection, since the direct product of N=4 Yang-Mills (from the N=2 open string) with N=4 Yang-Mills (from the dimensional reduction of the D=10 N=1 open string) is N=8 supergravity. Self-dual extended (N>3) gauged supergravities avoid some of the awkward features of the corresponding non-self-dual theories, in particlular the nonlinear scalar potential and the cosmological constant.

The amplitudes that follow from our supersymmetric, manifestly Lorentz invariant actions for (uncoupled) open and closed strings, when written in the string gauge, are the same as those obtained from string calculations [2,4]. (The string gauge for those actions [9] is closely related to the light-cone gauge [10].) Those for the heterotic string are the same at least when restricted to Yang-Mills interactions. (We have not compared the heterotic gravitational couplings because those of ref. [3] have been given in a different gauge from all the other amplitudes.) Although the vanishing of loop corrections in the open and closed strings follows directly from the absence of fermionic derivatives in the action, the heterotic case is again more subtle: Fermionic derivatives appear, but few enough so that loops with fewer than four legs vanish automatically. The rest (four external legs or more) may then vanish for kinematic reasons, as for tree graphs [11,2-4].

## 2. The self-dual light-cone

We first consider general properties of the light-cone gauge for self-dual theories, which is simpler then the string gauge. Later we'll analyze the covariant action, from which both the light-cone and string gauges can be chosen.

The light-cone formalism for self-dual theories is simpler than the usual light-cone formalism because more of the Lorentz symmetry is manifest: The coordinates $x^{\alpha\alpha'}$ are a representation of SO(2,2)=SL(2)⊗SL(2)′, where only SL(2)′ is broken, down to GL(1). Then $x^{\alpha-'}$ is treated as "time" coordinates and $x^{\alpha+'}$ as "space" coordinates, although both are light-like (as $x^-$ and $x^+$ in the usual light cone). The kinetic operator $\Box = i\partial^\alpha{}_{-'}\partial_{\alpha+'}$ is linear in time derivatives. The interaction terms must contain only space derivatives $\partial_{\alpha+'}$ and no time derivatives $\partial_{\alpha-'}$. The manifest GL(2) invariance forbids the $1/\partial_{++'}$'s of the usual light-cone formalism, so the action is local. For the theories considered here, we find only cubic interaction terms, as for the almost identical (but noncovariant) actions proposed earlier for self-dual Yang-Mills and gravity [10].

Gauge fixing is similar to the usual light cone, but taken a step further: By choosing a light-cone gauge, eliminating auxiliary degrees of freedom, and imposing self-duality, all field strengths $F$ and gauge fields $A$ can be expressed in terms of "prepotentials" $V$ with just a single component, corresponding to the single helicity it represents:

$$F_{\alpha_1...\alpha_{2s}} = -i\partial_{\alpha_1+'}...\partial_{\alpha_{[s]}+'}A_{\alpha_{[s+1]}...\alpha_{2s}-'..._-'} = \partial_{\alpha_1+'}...\partial_{\alpha_{2s}+'}V_{-'..._-'}$$

(Its eigenvalue of the GL(1) generator, represented by the number of $-'$ indices, is just twice the helicity $s$. "[$s$]" here means the greatest integer in $s$.) Such prepotentials are known from supersymmetric theories, where light-cone gauges induce similar relations. This expression is also correct for the interacting case, since we use only space derivatives, and the gauge fields contain only time components in this gauge. There are also Lagrange multiplier field strengths $F_{\alpha'_1...\alpha'_{2s}}$ (unrelated to $F_{\alpha_1...\alpha_{2s}}$, with different gauge fields), describing negative helicity: In the light cone these appear only as $F_{+'...+'}$, since the other components do not generate equations of motion whose solution requires inverting time derivatives (i.e. they are auxiliary fields).

We'll show below that such relations generalize in a trivial way to supersymmetric self-dual theories, by generalizing the SL(2) index $\alpha$ to an SL(N|2) index $A = (a, \alpha)$ on both the fields and the coordinates, thus introducing anticommuting coordinates:

$$F_{A_1...A_{2s}} = -i\partial_{A_1+'}...\partial_{A_{2s}+'}V_{-'...-'}$$

where $s$ is now the *maximum* helicity of the supersymmetric multiplet. (Lagrange multiplier field strengths come from hitting it with more $\partial_{a+'}$'s.) The free field equation $\Box V = 0$ for the prepotential also generalizes:

$$\partial_A{}^{\alpha'}\partial_{B\alpha'}V_{-'...-'} = 0$$

There is a similar generalization for the free action. Breaking the SL(N|2) covariance by solving this field equation for all $\theta^{a-'}$ dependence, $V$ is reduced to a superfield that depends on just $x^{\alpha\alpha'}$ and $\theta^{a+'}$:

$$S_0 = \int d^4x\, d^N\theta\, \tfrac{1}{2}V\Box V$$

By dimensional analysis, $-4+\tfrac{1}{2}N+2(1-s)+2 = 0 \Rightarrow N = 4s$, which implies maximal supersymmetry. We also have the same result from GL(1): $-N+2(2s) = 0$. Thus, the condition that all fields are in the same supersymmetric multiplet requires maximal supersymmetry. (In particular, the similar N=0 actions proposed for non-super self-dual Yang-Mills in the light-cone gauge and the string gauge are not only Lorentz non-covariant but are not even consistent with respect to dimensional analysis.)

## 3. Self-dual superspace

Self-dual super Yang-Mills is described by the commutation relations of the gauge-covariant superspace derivatives:

$$\{\nabla^a{}_\alpha, \nabla^b{}_\beta\} = 0, \quad \{\nabla^a{}_\alpha, \nabla_{b\beta'}\} = \delta^a_b \nabla_{\alpha\beta'}, \quad [\nabla^a{}_\alpha, \nabla_{\beta\beta'}] = 0$$

$$\{\nabla_{a\alpha'}, \nabla_{b\beta'}\} = C_{\beta'\alpha'}\phi_{ab}, \quad [\nabla_{a\alpha'}, \nabla_{\beta\beta'}] = C_{\beta'\alpha'}\chi_{a\beta}, \quad [\nabla_{\alpha\alpha'}, \nabla_{\beta\beta'}] = C_{\beta'\alpha'}F_{\alpha\beta}$$

The commutation relations give the field equations as well as the supersymmetry transformations. The latter set of commutators can be written more succinctly as

$$[\nabla_{A\alpha'}, \nabla_{B\beta'}\} = C_{\beta'\alpha'}F_{AB}$$

in terms of the indices $A, B$ of SL(N|2), which is a global symmetry of the commutation relations.

The generalization of the former set of commutation relations for gauged self-dual supergravity is

$$\{\nabla^{a\alpha}, \nabla^{b\beta}\} = C^{\alpha\beta}M^{ab} + \eta^{ab}M^{\alpha\beta}$$

$$\{\nabla^{a\alpha}, \nabla_{b\beta'}\} = \delta^a_b C^{\alpha\beta}\nabla_{\beta\beta'}, \quad [\nabla^{a\alpha}, \nabla_{\beta\beta'}] = \delta^\alpha_\beta \eta^{ab}\nabla_{b\beta'}$$

where $M^{ab}$ are the gauge generators of the gauge group SO(N) of the vectors, or more generally SO(n,N-n) or their contractions. Contraction to U(1)$^{N(N-1)/2}$ gives ungauged supergravity. Both $M^{ab}$ and the corresponding group metric $\eta^{ab}$ have dimensions of mass: This metric has eigenvalues $0, \pm m$, where $m = g/\kappa$ is the ratio of gauge and gravitational couplings. (The inverse metric $\eta_{ab}$ never appears in the self-dual theory, and doesn't even exist for the group contractions.) $M_{\alpha\beta}$ is the self-dual half of the local Lorentz generators. These constraints (and those below) follow uniquely from the self-dual restriction of the non-self-dual theory, except that the reality condition on the scalars for N=8 is changed: Instead of restricting the "self-dual" scalars to be the complex conugate of the "anti-self-dual" scalars, the anti-self-dual scalars are set to vanish [5]. (For background on the non-self-dual case, see [12] for the ungauged theories, [13] for the gauged N=4 theory, and [14] for the gauged N=8 theory in the equivalent component formalism. Since non-self-dual N=8 supergravity is much more complicated than the self-dual case, even in superspace, we do not reproduce it here. Self-duality for supergravity for N<4, where there are no scalars, was first discussed in [15]; Wick rotation of those theories to 2+2 dimensions is trivial.)

From these commutation relations we can recognize that (1) $M^{ab}$, $M^{\alpha\beta}$, and $\nabla^{a\alpha}$ are the generators $M^{AB}$ of the algebra OSp(N|2) (or a contraction, etc.), with metric $\eta^{AB} = (\eta^{ab}, C^{\alpha\beta})$, and (2) $\nabla_{A\alpha'}$ is in the defining representation. (In the Yang-Mills case, $\nabla^{a\alpha}$ is half of the fermionic generators of the manifest global SL(N|2) symmetry, a subgroup of the non-manifest (S)SL(N|4) superconformal symmetry, which is here broken to the OSp(N|2) subgroup.) We therefore choose to interpret $\nabla^{a\alpha}$ as a local symmetry generator rather than as a covariant coordinate derivative. Our superspace coordinates are then just $x^{M\mu'} = (x^{\mu\mu'}, \theta^{m\mu'})$. Furthermore, since all field strengths of the local SL(2)$'$ and SL(N) generators vanish anyway, we can drop those generators from the local symmetry group, which is now just OSp(N|2) (plus general coordinate transformations in (N|2)+(N|2) dimensions). The covariant derivatives thus take the form

$$\nabla_{A\alpha'} = E_{A\alpha'}{}^{M\mu'}\partial_{M\mu'} + \tfrac{1}{2}\Omega_{A\alpha' BC}M^{CB}$$

(As usual, there are extra implicit sign factors from reordering of fermionic indices.) Our conventions for OSp(N|2) are that on the fundamental representation $[M^{AB}, V^C\} = V^{[A}\eta^{B)C}$; we then have the useful identities

$$\tfrac{1}{2}K_{BC}[M^{CB}, V^A\} = -\eta^{AB}V^C K_{CB}, \quad \tfrac{1}{2}K_{BC}[M^{CB}, V_A\} = K_{AB}\eta^{CB}V_C$$

These determine the other representations and the commutation relations of the generators themselves. The grading is defined by treating $\alpha$ and $\alpha'$ as bosonic indices, and $a$ as fermionic; $\eta$ is graded antisymmetric in its indices, while $M$ is graded symmetric. In analogy to the Yang-Mills case, the commutation relations of the covariant derivatives are

$$[\nabla_{A\alpha'}, \nabla_{B\beta'}\} = C_{\beta'\alpha'}\tfrac{1}{2}F_{ABCD}M^{DC}$$

where $F_{ABCD}$ is totally graded symmetric, and at $\theta = 0$ gives the nonnegative-helicity field strengths. (There is no analog to the terms in $\{\nabla^{a\alpha}, \nabla^{b\beta}\}$, so the superspace is in some sense "half DeSitter.") Yang-Mills can be treated in the same superspace, with $\nabla_{A\alpha'} = \partial_{A\alpha'} + A_{A\alpha'}$. (Yang-Mills is superconformal, so how SL(N|2) is broken for that theory doesn't matter.) The solution to these relations can now be treated in exactly the same way as the nonsupersymmetric case, simply by generalizing the indices from SL(2)⊗SL(2)' to OSP(N|2)⊗SL(2)' (local⊗global).

The field equations, supersymmetry transformations, and expressions for the field strengths then follow from these commutators and their Bianchi identities

$$\nabla_{[A\alpha'}F_{B)CDE} = 0$$

and their derivatives. We then find the series of identities

$$\nabla_{A\alpha'}F_{BCDE} = F_{ABCDE\alpha'}$$

$$\nabla_{A\alpha'}F_{BCDEF\beta'} = F_{ABCDEF\alpha'\beta'} + C_{\beta'\alpha'}\tfrac{1}{24}F_{A(BC|G}\eta^{HG}F_{H|DEF]}$$

and identities of the form $\nabla F'' = F''' + FF'$, $\nabla F''' = F'''' + FF'' + F'^2$, and $\nabla F'''' = F''''' + FF''' + F'F''$, where all the $F$'s are totally symmetric in their SL(2)' indices and totally graded symmetric in their OSp(N|2) indices. At $\theta = 0$ with all OSp(N|2) indices restricted to SO(N) indices, they are the nonpositive-helicity field strengths.

The self-dual theory is considerably simpler than the non-self-dual one, avoiding nonlinearities in the scalar fields. This theory also has no cosmological constant, unlike the general result for the non-self-dual case in the absence of spontaneous supersymmetry breakdown [16], since the cosmological constant would appear as $\eta^{ab}\eta_{ab}$, whereas in the self-dual theory only $\eta^{ab}$ ever appears anywhere. (This is actually a consequence of the fact that even in the non-self-dual theories the SO(N) internal group metrics $\eta^{ab}$ and $\eta_{ab}$ which appear as complex conjugates in 3+1 dimensions are independent in 2+2 dimensions.)

Another example of self-dual supersymmetric theories is the scalar multiplet: The free case is described by
$$\partial_{[A\alpha'}F_{B)b'} = 0$$
where $b'$ is an internal symmetry index.

## 4. Light-cone action

One way to analyze the commutation relations is by going to the light-cone gauge, where half of the Lorentz symmetry is manifest and half of the supersymmetry. We first separate the constraints by breaking SL(2)':

$$[\nabla_{A+'}, \nabla_{B+'}\} = 0$$

$$[\nabla_{A(+'}, \nabla_{B-')}\} = 0$$

$$[\nabla_A{}^{\alpha'}, \nabla_{B\alpha'}\} = F_{ABCD}M^{DC}$$

$$[\nabla_{A-'}, \nabla_{B-'}\} = 0$$

(All except the third are the same as in Yang-Mills.) The first three constraints can be solved explicitly:

$$\nabla_{A+'} = \partial_{A+'}$$

$$\Delta_{-'-'} = 2(\partial_{A+'}V_{-'-'-'-'})\eta^{BA}\partial_{B+'} + \tfrac{1}{2}(\partial_{A+'}\partial_{B+'}V_{-'-'-'-'})M^{BA}$$

$$\nabla_{A-'} = \partial_{A-'} + [\partial_{A+'}, \Delta_{-'-'}]$$
$$= \partial_{A-'} + (\partial_{A+'}\partial_{B+'}V_{-'-'-'-'})\eta^{CB}\partial_{C+'} + \tfrac{1}{2}(\partial_{A+'}\partial_{B+'}\partial_{C+'}V_{-'-'-'-'})M^{CB}$$

$$F_{ABCD} = -i\partial_{A+'}\partial_{B+'}\partial_{C+'}\partial_{D+'}V_{-'-'-'-'}$$

The remaining constraint reduces to:
$$\partial_A{}^{\alpha'}\partial_{B\alpha'}V_{-'-'-'-'} + i(\partial_{A+'}\partial_{C+'}V_{-'-'-'-'})\eta^{DC}(\partial_{D+'}\partial_{B+'}V_{-'-'-'-'}) = 0$$

For comparison, Yang-Mills gives

$$\nabla_{A+'} = \partial_{A+'}$$

$$\nabla_{A-'} = \partial_{A-'} + (\partial_{A+'}V_{-'-'})$$

$$F_{AB} = -i\partial_{A+'}\partial_{B+'}V_{-'-'}$$

$$\partial_A{}^{\alpha'}\partial_{B\alpha'}V_{-'-'} + i[(\partial_{A+'}V_{-'-'}), (\partial_{B+'}V_{-'-'})\} = 0$$

while for the free scalar multiplet

$$F_{Aa'} = -i\partial_{A+'}V_{-'a'}$$

$$\partial_A{}^{\alpha'}\partial_{B\alpha'}V_{-'a'} = 0$$

The string gauge is related to the light-cone gauge (see [10,5] for details): There the roles of solved constraint and field equation are switched between $[\nabla_{A(+'}, \nabla_{B-')}\} = 0$ and $[\nabla_{A-'}, \nabla_{B-'}\} = 0$. As a result, although the free term of the field equation (action) is the same, the interaction is more complicated (nonpolynomial in the Yang-Mills and gauged supergravity cases) and contains time derivatives.

The parts involving $\partial_{a-'}$ can be solved for all $\theta^{a-'}$ dependence, so $V$ can be taken as evaluated at $\theta^{a-'} = 0$. This leaves just the $\alpha\beta$-part of the last constraint as the equation of motion: For supergravity,

$$\Box V + \tfrac{1}{2}i(\partial^\alpha{}_{+'}\partial_{B+'}V)\eta^{CB}(\partial_{C+'}\partial_{\alpha+'}V) = 0$$

The action is then

$$S_8 = \int d^4x\, d^8\theta\, \tfrac{1}{2}V\Box V + \tfrac{1}{6}iV(\partial^\alpha{}_{+'}\partial_{B+'}V)\eta^{CB}(\partial_{C+'}\partial_{\alpha+'}V)$$

This should be compared with the N=4 supersymmetric self-dual Yang-Mills action that follows from the open string [5], which differs from earlier noncovariant proposals [10] only by the appearance of $\theta^{a+'}$ as coordinates:

$$S_4 = \int d^4x\, d^4\theta\, \tfrac{1}{2}V\Box V + \tfrac{1}{3}iV(\partial^\alpha{}_{+'}V)(\partial_{\alpha+'}V)$$

(A similar N=3, and therefore noncovariant, action follows from the self-dual restriction of the light-cone action of [17].) The supergravity interaction has two terms, from the $\eta^{CB}$: the Yang-Mills-type interaction ($\eta^{cb}$) term for gauged supergravity (heterotic string) and the gravitational-type ($C^{\gamma\beta}$) term for both gauged and ungauged (closed string). The $\eta^{cb}$ term is lower in spacetime derivatives because of the dimensional $g/\kappa$ coupling, but contains $\theta$ derivatives, unlike the nonheterotic cases. It has the same momentum dependence as the interaction in the supersymmetric Yang-Mills theory. The universal (minimal) couplings of the gluon of the open string and the graviton of the closed string follow from the fact they appear in their respective $V$'s at $\theta = 0$, so they have contributions to the action which are $\theta$ (spin) independent.

As for self-dual Yang-Mills [5], ungauged self-dual supergravity has all loop graphs vanishing because of the absence of spinor derivatives in the action. (These cases are similar to the $\mathbf{Z}_n$ model of [2], which also describes a multiplet of various spins, but here there are fermions and $V$ is a true superfield. There is no decoupling of fields in the "diagonalized" field $V(\theta)$ because it cannot be interpreted as independent fields for different "values" of $\theta$.) However, this is not the case for gauged supergravity, where the appearance of spinor derivatives for the Yang-Mills coupling requires a graph-by-graph analysis. In terms of the number of vertices V, propagators P, external lines E, and loops L, the relation that all vertices are 3-point is 3V=2P+E, the Gauss-Bonet theorem ($\hbar$ counting) is L−1=P−V, and the condition that each loop needs at least 8 $\theta$ derivatives to kill each $\int d^8\theta$ [18] is 2V≥8L. These imply E≥2(L+1), so graphs with fewer than 2(L+1) external lines vanish. Since tree graphs vanish for E>3 by another mechanism [2-4], the other loop graphs may vanish for the same reason; in any case, the 3-point graph gets no quantum corrections.

## 5. Covariant field equations for helicity>1

Another useful gauge is the Wess-Zumino gauge, which is Lorentz covariant but has no manifest supersymmetry. (See [19] for more detail and a general discussion.) The gauge transformations $\nabla' = e^{-K}\nabla e^K$ for the gauge fields follow from commutators with gauge generators

$$K = K^{A\alpha'}\nabla_{A\alpha'} + \tfrac{1}{2}K_{AB}M^{BA}$$
$$= \zeta^{\alpha\alpha'}\nabla_{\alpha\alpha'} + \epsilon^{a\alpha'}\nabla_{a\alpha'} + \tfrac{1}{2}\lambda_{\alpha\beta}M^{\alpha\beta} + \epsilon_{a\alpha}M^{a\alpha} + \tfrac{1}{2}\xi_{ab}M^{ba}$$

All gauge fields (for positive helicity) are found in the vector covariant derivative at $\theta = 0$:

$$\nabla_{\alpha\alpha'}|_{\theta=0} = e_{\alpha\alpha'}{}^{\mu\mu'}\partial_{\mu\mu'} + \psi_{\alpha\alpha'}{}^{m\mu'}\partial_{m\mu'} + \tfrac{1}{2}\omega_{\alpha\alpha'\beta\gamma}M^{\beta\gamma} + \psi_{\alpha\alpha'b\gamma}M^{b\gamma} + \tfrac{1}{2}A_{\alpha\alpha'bc}M^{cb}$$

Part of the Wess-Zumino gauge choice is to use the $\theta$-coordinate transformations to choose

$$\nabla_{a\alpha'}|_{\theta=0} = \partial_{a\alpha'} \quad \Rightarrow \quad E_{A\alpha'}{}^{M\mu'}|_{\theta=0} = \begin{pmatrix} e_{\alpha\alpha'}{}^{\mu\mu'} & \psi_{\alpha\alpha'}{}^{m\mu'} \\ 0 & \delta_a^m\delta_{\alpha'}^{\mu'} \end{pmatrix}$$

$$\Rightarrow \quad E_{M\mu'}{}^{A\alpha'}|_{\theta=0} = \begin{pmatrix} e_{\mu\mu'}{}^{\alpha\alpha'} & -e_{\mu\mu'}{}^{\beta\beta'}\psi_{\beta\beta'}{}^{n\nu'}\delta_n^a\delta_{\nu'}^{\alpha'} \\ 0 & \delta_m^a\delta_{\mu'}^{\alpha'} \end{pmatrix}$$

where $E_{M\mu'}{}^{A\alpha'}$ is the inverse of $E_{A\alpha'}{}^{M\mu'}$. Similarly, gauge choices can be made for terms in $\nabla$ linear in $\theta$ such that they include only these gauge fields and the field strengths $F_{ABCD}|_{\theta=0}$.

In this gauge, some component equations (specifically, field equations for gauge fields) are simpler when expressed in Cartan notation, where a mixture of curved and flat indices are used. Then the general form of the torsion and curvature (before imposing any constraints) is

$$[\nabla_\mathbf{A}, \nabla_\mathbf{B}\} = T_{\mathbf{AB}}{}^\mathbf{C}\nabla_\mathbf{C} + \tfrac{1}{2}R_{\mathbf{AB}CD}M^{DC}$$

$$\begin{aligned}\Rightarrow \quad T_{\mathbf{MN}}{}^\mathbf{A} &= -\partial_{[\mathbf{M}}E_{\mathbf{N})}{}^\mathbf{A} + E_{[\mathbf{M}}{}^{B\alpha'}\Omega_{\mathbf{N})BC}\eta^{AC} \\ R_{\mathbf{MN}AB} &= \partial_{[\mathbf{M}}\Omega_{\mathbf{N})AB} - \Omega_{[\mathbf{M}AC}\eta^{DC}\Omega_{\mathbf{N})DB}\end{aligned}$$

where $\mathbf{A} = A\alpha'$, $\mathbf{M} = M\mu'$. The corresponding form of the gauge transformation laws is

$$\delta\nabla_\mathbf{A} = [\nabla_\mathbf{A}, K^\mathbf{B}\nabla_\mathbf{B} + \tfrac{1}{2}K_{BC}M^{CB}]$$

$$\begin{aligned}\Rightarrow \quad \delta E_\mathbf{M}{}^\mathbf{A} &= -\nabla_\mathbf{M}K^\mathbf{A} + K^\mathbf{N}T_{\mathbf{NM}}{}^\mathbf{A} + \eta^{AB}E_\mathbf{M}{}^{C\alpha'}K_{CB} \\ \delta\Omega_{\mathbf{M}AB} &= -K^\mathbf{N}R_{\mathbf{NM}AB} + \nabla_\mathbf{M}K_{AB} - \eta^{DC}K_{(A|C}\Omega_{\mathbf{M}D|B)}\end{aligned}$$

Conversion of these curved indices back to flat indices is then easy at $\theta = 0$, where for a general covariant supervector

$$V_{a\alpha'} = \delta_a^m\delta_{\alpha'}^{\mu'}V_{m\mu'}, \quad V_{\alpha\alpha'} = e_{\alpha\alpha'}{}^{\mu\mu'}V_{\mu\mu'} + \psi_{\alpha\alpha'}{}^{b\beta'}V_{b\beta'}$$

Therefore, from now on we use flat and curved spinor super-indices $a\alpha'$ and $m\mu'$ interchangeably, while flat and curved vector indices on component fields are converted with the vierbein $e_{\alpha\alpha'}{}^{\mu\mu'}$, with the $\psi_{\alpha\alpha'}{}^{b\beta'}$ terms explicit.

The only ambiguities in finding an action principle for these equations are which fields are to be taken as independent, and correspondingly which equations are to be taken as equations of motion which follow from the action (as opposed to constraints which are imposed in defining the field content). To treat both the gauged and ungauged theories in the same formalism, we'll need to use a first-order formalism for the gravitino, analogous to that frequently used for the graviton. In ordinary (super)gravity, a torsion constraint determines the Lorentz connection $\omega$ in terms of the vierbein $e$; here, another torsion constraint detemines the left-handed gravitino field $\psi_{\alpha\alpha'b\beta}$ in terms of the right-handed gravitino field $\psi_{\alpha\alpha'}{}^{m\mu'}$. (Both gravitino fields here describe helicity $+3/2$.) In ordinary supergravity, a torsion constraint gives the field equation for the gravitino; here, another torsion constraint gives the field equation for the graviton. (This is possible because SL(2)$'$ is not gauged, unlike ordinary gravity.) Thus, for gauged supergravity, all these constraints and field equations for the graviton and gravitino are just the complete set of vector-vector torsions evaluated at $\theta = 0$. Separating out the graviton and gravitino equations,

$$-T_{\mathbf{mn}}{}^{\alpha\alpha'} = \partial_{[\mathbf{m}} e_{\mathbf{n}]}{}^{\alpha\alpha'} + e_{[\mathbf{m}}{}^{\beta\alpha'}\omega_{\mathbf{n}]\beta}{}^{\alpha} + \psi_{[\mathbf{m}}{}^{b\alpha'}\psi_{\mathbf{n}]b}{}^{\alpha} = 0$$

$$-T_{\mathbf{mn}}{}^{a\alpha'} = \partial_{[\mathbf{m}} \psi_{\mathbf{n}]}{}^{a\alpha'} + \eta^{ab}\psi_{[\mathbf{m}}{}^{c\alpha'}A_{\mathbf{n}]bc} - \eta^{ab}e_{[\mathbf{m}}{}^{\alpha\alpha'}\psi_{\mathbf{n}]b\alpha} = 0$$

(The bold indices are curved vector indices, $\mathbf{m} = \mu\mu'$.) A Lorentz decomposition of these equations (after using the vierbein $e_{\alpha\alpha'}{}^{\mathbf{m}}$ to flatten all curved indices) shows that some parts determine $\omega$ and $\psi_{\alpha\alpha'b\beta}$ completely. The remaining parts are the field equations for $e$ and $\psi_{\alpha\alpha'}{}^{b\beta'}$: A light-cone gauge analysis reproduces the results obtained earlier.

To include the gauge-contracted theories, we include the usual field equation for $\psi_{\alpha\alpha'b\beta}$. This is redundant in gauged supergravity. Its simplest form is obtained by taking the curl of the torsion equation and factoring out an $\eta^{ab}$:

$$\partial_{[\mathbf{m}} E_{\mathbf{n}}{}^{B\alpha'}\Omega_{\mathbf{p}]Ba} = \partial_{[\mathbf{m}}(e_{\mathbf{n}}{}^{\beta\alpha'}\psi_{\mathbf{p}]a\beta} - \psi_{\mathbf{n}}{}^{b\alpha'}A_{\mathbf{p}]ba}) = 0$$

$$\tfrac{1}{2}\partial_{[\mathbf{M}}T_{\mathbf{NP})}{}^{\mathbf{A}} = \eta^{AC}(\partial_{[\mathbf{M}}E_{\mathbf{N}}{}^{B\alpha'}\Omega_{\mathbf{P})BC})$$

More covariant, but more complicated, forms of this equation can be obtained from the covariant curl of this torsion, or from the curvature $\tfrac{1}{2}R_{[\mathbf{mn}aB}e_{\mathbf{p}]}{}^{B\alpha'}$ (which is the covariant curl of the torsion plus torsion squared terms with an $\eta^{ab}$ factored out, according to the Bianchi identities). This equation is similar to the $\omega$-independent part of the graviton equation:

$$-\tfrac{1}{2}T_{[\mathbf{mn}}{}^{\alpha(\alpha'}e_{\mathbf{p}]\alpha}{}^{\beta')} = \tfrac{1}{2}\partial_{[\mathbf{m}}e_{\mathbf{n}}{}^{\alpha(\alpha'}e_{\mathbf{p}]\alpha}{}^{\beta')} + \psi_{[\mathbf{m}}{}^{a(\alpha'}\psi_{\mathbf{n}a}{}^{\alpha}e_{\mathbf{p}]\alpha}{}^{\beta')} = 0$$

On the other hand, in ungauged supergravity the gravitino's torsion constraint simply says that $\psi_{\alpha\alpha'}{}^{m\mu'}$ is pure gauge. (The intermediate group contractions have mixtures of these conditions, some components of each chirality of $\psi$ being trivial.)

The field equation for $e$ appears with one more derivative in the covariant lagrangian than in the light-cone one: The light-cone equations of motion $f_{AB}$, of which the part $\frac{1}{2}f^\alpha{}_\alpha = \Box V + ...$ follows from the light-cone superfield action, actually appears in the covariant derivative commutators as

$$i[\nabla_{A-'}, \nabla_{B-'}\} = (\partial_{C+'} f_{AB})\eta^{DC}\partial_{D+'} + \tfrac{1}{2}(\partial_{C+'}\partial_{D+'} f_{AB})M^{DC}$$

It is part of the first term, which is a torsion constraint, which follows from the covariant lagrangian.

## 6. Lorentz covariant action

We have not yet completed a full analysis of the Lorentz covariant component action. However, the results given in the previous sections are sufficient to obtain the action itself, since the field equations for positive helicity are imposed by Lagrange multipliers; i.e. the fields describing negative helicity appear only linearly, so their field equations and the complete action are implied by the field equations of the other fields. Specifically, the field equations for helicity>1 were obtained by analyzing torsion constraints contained in the commutation relations in the previous section; the field equation for helicity 1 is simply the self-duality of the (supercovariant) Yang-Mills field strength (also implied by the commutation relations); that for helicity 1/2 is contained in the Bianchi identity $\nabla F = F'$ of section 3; and that for 0 (and $-1/2$) is contained in the $\nabla F' = F'' + F^2$ identity of that section. The remaining field equations can also be obtained from the rest of that series of identities. The supersymmetry transformations can be obtained by a similar analysis: Those for helicity>1/2 follow directly from the commutation relations, as described in the previous section; those for helicities 1/2 and 0 follow from the Bianchi identity; those for the negative helicities follow from the rest of that series of identities.

The Lorentz covariant lagrangian is then found to be

$$\begin{aligned}L =& \epsilon^{\mathbf{mnpq}}(-\tfrac{1}{2}\tilde{\omega}_{\mathbf{m}}{}^{\alpha'\beta'} T_{\mathbf{np}}{}^\alpha{}_{\alpha'} e_{\mathbf{q}\alpha\beta'} + \tilde{\psi}_{\mathbf{m}}{}^{a\alpha'} \partial_{\mathbf{n}} E_{\mathbf{p}}{}^B{}_{\alpha'}\Omega_{\mathbf{q}Ba} - \tfrac{1}{4}\widetilde{R}_{\mathbf{mna}}{}^{\alpha'} T_{\mathbf{pq}}{}^a{}_{\alpha'}) \\ &+ \mathbf{e}^{-1}[\tfrac{1}{4}G^{ab\alpha'\beta'} F_{ab\alpha'\beta'} + \tfrac{1}{6}\chi^{abc\alpha'}\nabla^\alpha{}_{\alpha'}\chi_{abc\alpha} - \tfrac{1}{2304}\epsilon^{abcdefgh}(\nabla^{\alpha\alpha'}\phi_{abcd})(\nabla_{\alpha\alpha'}\phi_{efgh}) \\ &- \epsilon^{abcdefgh}(\tfrac{1}{192}\phi_{abcd}F_{ef}{}^{\alpha\beta}F_{gh\alpha\beta} - \tfrac{1}{72}\chi_{abc}{}^\alpha\chi_{def}{}^\beta F_{gh\alpha\beta} + \tfrac{1}{192}\eta^{ij}\phi_{abcd}\chi_{efi}{}^\alpha\chi_{ghj\alpha})]\end{aligned}$$

where

$$\nabla_{\alpha\alpha'}\phi_{abcd} = D_{\alpha\alpha'}\phi_{abcd} - \tfrac{1}{6}\psi_{\alpha\alpha'[a}{}^\beta \chi_{bcd]\beta} + \psi_{\alpha\alpha'}{}^{e\beta'}\chi_{abcde\beta'}$$

$$\nabla_{\alpha\alpha'}\chi_{abc\beta} = D_{\alpha\alpha'}\chi_{abc\beta} + \psi_{\alpha\alpha'd\gamma}(\delta^\gamma_\beta \eta^{de}\phi_{eabc} + \tfrac{1}{2}C^{\delta\gamma}\delta^d_{[a}F_{bc]\delta\beta}) + \psi_{\alpha\alpha'}{}^{d\beta'}\nabla_{\beta\beta'}\phi_{abcd}$$

$$F_{\mathbf{mn}ab} = f_{\mathbf{mn}ab} - \psi_{[\mathbf{m}a}{}^\gamma \psi_{\mathbf{n}]b\gamma} + \psi_{[\mathbf{m}}{}^{c\alpha'}(e_{\mathbf{n}]}{}^\alpha{}_{\alpha'}\chi_{abc\alpha} + \tfrac{1}{2}\psi_{\mathbf{n}]}{}^d{}_{\alpha'}\phi_{abcd})$$

$$F_{ab\alpha\beta} = e_\alpha{}^{\alpha'\mathbf{m}} e_{\beta\alpha'}{}^{\mathbf{n}} F_{\mathbf{mn}ab}, \quad F_{ab\alpha'\beta'} = e^\alpha{}_{\alpha'}{}^{\mathbf{m}} e_{\alpha\beta'}{}^{\mathbf{n}} F_{\mathbf{mn}ab}$$

and $f$ is the Yang-Mills field strength found from $D_{\alpha\alpha'}$, the vector derivative covariant with respect to just Yang-Mills and gravity (without gauging SL(2)$'$). The independent fields are $e$ (helicity $+2$), $\psi$ and $\psi'$ ($+3/2$), $A$ ($+1$), $\chi$ ($+1/2$), $\phi$ (0), $\chi'$ ($-1/2$), $G$ ($-1$), $\tilde\psi$ and $\widetilde R$ ($-3/2$), and $\tilde\omega$ ($-2$). (If we had used a first-order formalism for the graviton, the full $T_{\mathbf{mn}}{}^{\alpha\alpha'}$ constraint would appear, with $\omega_{\mathbf{m}\alpha\beta}$ also an independent field, and $\tilde\omega$ part of a Lagrange multiplier $\widetilde R_{\mathbf{mn}\alpha\alpha'}$.) Although all negative helicity fields appear only as Lagrange multipliers, the kinetic terms $\tilde\psi\psi$, $\chi'\chi$, and $\phi\phi$ have the usual form. All fields with upper SL(8) indices (except $\psi_{\mathbf{m}}{}^{a\alpha'}$) have been obtained from those with just lower indices by using $\epsilon^{abcdefgh}$ (including $\tilde\omega$, which originally had 8 lower indices): In terms of the original fields, each term in $L$ (except the $\widetilde R$ term) has exactly one $\epsilon^{abcdefgh}$ factor. (This means that the action has a specific quantum number of the GL(1) subgroup of the GL(8) invariance of the ungauged field equations.) $G$ and $\tilde\omega$ appear as in the first-order formalism for spins 1 and 2, but they have no quadratic terms, and they are anti-self-dual.

$\tilde\psi$, $\widetilde R$, and $\tilde\omega$ are abelian gauge fields which don't appear explicitly in $\nabla$ at $\theta = 0$; their on-shell field strengths appear in the $\theta^{a\alpha'}$ expansion of

$$\begin{aligned}F_{abcd} =& \phi_{abcd} + \theta^{e\alpha'}\chi_{abcde\alpha'} + \tfrac{1}{2}\theta^{e\alpha'}\theta^{f\beta'} G_{abcdef\alpha'\beta'} \\ &+ \tfrac{1}{6}\theta^{e\alpha'}\theta^{f\beta'}\theta^{g\gamma'} \widetilde R_{abcdefg\alpha'\beta'\gamma'} + \tfrac{1}{24}\theta^{e\alpha'}\theta^{f\beta'}\theta^{g\gamma'}\theta^{h\delta'} \widetilde W_{abcdefgh\alpha'\beta'\gamma'\delta'} + ...\end{aligned}$$

as

$$\begin{aligned}\widetilde R^a{}_{\alpha'\beta'\gamma'} &= e^\alpha{}_{(\alpha'}{}^{\mathbf{m}} e_{\alpha\beta'}{}^{\mathbf{n}}(\partial_{\mathbf{m}}\tilde\psi^a{}_{\mathbf{n}\gamma')} - \tilde\omega_{\mathbf{m}\gamma')}{}^{\delta'}\psi^a_{\mathbf{n}\delta'} - \tfrac{1}{2}\eta^{ab}\widetilde R_{\mathbf{mn}b\gamma')}) \\ \widetilde W_{\alpha'\beta'\gamma'\delta'} &= e^\alpha{}_{(\alpha'}{}^{\mathbf{m}} e_{\alpha\beta'}{}^{\mathbf{n}} \partial_{\mathbf{m}}\tilde\omega_{\mathbf{n}\gamma'\delta')}\end{aligned}$$

These field strengths appear at $\theta = 0$ in the covariant quantitites $F_{ABCEDFG\alpha'\beta'\gamma'}$ and $F_{ABCEDFGH\alpha'\beta'\gamma'\delta'}$ (and $\chi_{abcde\alpha'}$ and $G_{abcdef\alpha'\beta'}$ in $F_{ABCDE\alpha'}$ and $F_{ABCDEF\alpha'\beta'}$), which arise in the series of identities following from the Bianchi identities discussed in section 3. We also have, at $\theta = 0$, $F_{ABCD} = (W_{\alpha\beta\gamma\delta}, R_{a\beta\gamma\delta}, F_{ab\gamma\delta}, \chi_{abc\delta}, \phi_{abcd})$, the on-shell field strengths for nonnegative helicities. In the gauged theory, $\tilde\psi$ and $\psi_{\mathbf{m}a\alpha}$ are purely gauge plus auxiliary, and propagating helicity $\pm 3/2$ can be described completely in terms of $\widetilde R$ and $\psi_{\mathbf{m}}{}^a{}_{\alpha'}$; in the ungauged case the opposite is true; for nontrivial group contractions we have components of each.

Of the local symmetries, we have the usual forms for general coordinate, local Lorentz ($M_{\alpha\beta}$), and SO(8) (or the contractions, etc.), except that because of the Yang-Mills noncovariance of the field equation from varying $\tilde{\psi}$, there is an extra term $\tilde{\psi}_{[\mathbf{m}}{}^b{}_{\alpha'}\partial_{\mathbf{n}]}\xi_{ba}$ in the Yang-Mills gauge transformation of $\widetilde{R}_{\mathbf{mn}a\alpha'}$. We have not yet derived the complete local supersymmetry transformations, but they are roughly of the form

$$\begin{aligned}
\delta e_{\mathbf{m}}{}^{\alpha\alpha'} &= \epsilon^{\alpha\alpha'}\psi_{\mathbf{m}a}{}^{\alpha} & &+\epsilon_a{}^{\alpha}\psi_{\mathbf{m}}{}^{a\alpha'} \\
\delta\psi_{\mathbf{m}}{}^{a}{}_{\alpha'} &= D_{\mathbf{m}}\epsilon^a{}_{\alpha'} & &+\eta^{ab}\epsilon_b{}^{\alpha}e_{\mathbf{m}\alpha}{}^{\alpha'} \\
\delta\psi_{\mathbf{m}a\alpha} &= \epsilon^{b\alpha'}(e_{\mathbf{m}}{}^{\beta}{}_{\alpha'}F_{ab\alpha\beta} + \psi_{\mathbf{m}}{}^c{}_{\alpha'}\chi_{abc\alpha}) & &+D_{\mathbf{m}}\epsilon_{a\alpha} \\
\delta A_{\mathbf{m}ab} &= \epsilon^{c\alpha'}(e_{\mathbf{m}}{}^{\alpha}{}_{\alpha'}\chi_{abc\alpha} + \psi_{\mathbf{m}}{}^d{}_{\alpha'}\phi_{abcd}) & &+\epsilon_{[a}{}^{\alpha}\psi_{\mathbf{m}b]\alpha} \\
\delta\chi_{abc\alpha} &= \epsilon^{d\alpha'}\nabla_{\alpha\alpha'}\phi_{abcd} & &+\tfrac{1}{2}\epsilon_{[a}{}^{\beta}F_{bc]\alpha\beta} + \epsilon_{d\alpha}\eta^{de}\phi_{abce} \\
\delta\phi_{abcd} &= \epsilon^{e\alpha'}\chi_{abcde\alpha'} & &+\tfrac{1}{6}\epsilon_{[a}{}^{\alpha}\chi_{bcd]\alpha} \\
\delta\chi_{abcde\alpha'} &= \epsilon^{f\beta'}G_{abcdef\alpha'\beta'} + \tfrac{1}{12}\epsilon^f{}_{\alpha'}\eta^{gh}\phi_{abcg}\phi_{defh} & &+\tfrac{1}{24}\epsilon_{[a}{}^{\alpha}\nabla_{\alpha\alpha'}\phi_{bcde]} \\
\delta G^{ab}{}_{\alpha'\beta'} &= \epsilon^{[a\gamma'}\widetilde{R}^{b]}{}_{\alpha'\beta'\gamma'} + \tfrac{1}{6}\epsilon^{[a}{}_{(\alpha'}\eta^{b]c}\chi^{def}{}_{\beta')}\phi_{cdef} & &+\epsilon_c{}^{\alpha}\nabla_{\alpha(\alpha'}\chi^{abc}{}_{\beta')} \\
\delta\tilde{\psi}_{\mathbf{m}}{}^a{}_{\alpha'} &= \epsilon^{a\beta'}\tilde{\omega}_{\mathbf{m}\alpha'\beta'} & &+\epsilon_b{}^{\alpha}e_{\mathbf{m}\alpha}{}^{\beta'}G^{ab}{}_{\alpha'\beta'} \\
\delta\widetilde{R}_{\mathbf{mn}a\alpha'} &= \epsilon^{b\beta'}G^{cd}{}_{\alpha'\beta'}\phi_{abcd}? & &+? \\
\delta\tilde{\omega}_{\mathbf{m}\alpha'\beta'} &= 0? & &+\epsilon_a{}^{\alpha}\nabla_{\alpha(\alpha'}\tilde{\psi}_{\mathbf{m}}{}^a{}_{\beta')}\ (+\eta\widetilde{R}?)
\end{aligned}$$

There are also the abelian gauge symmetries

$$\delta\tilde{\omega}_{\mathbf{m}\alpha'\beta'} = \partial_{\mathbf{m}}\lambda_{\alpha'\beta'},\ \delta\tilde{\psi}_{\mathbf{m}}{}^a{}_{\alpha'} = \lambda_{\alpha'}{}^{\beta'}\psi_{\mathbf{m}}{}^a{}_{\beta'},\ \delta\widetilde{R}_{\mathbf{mn}a\alpha'} = \lambda_{\alpha'}{}^{\beta'}(e_{[\mathbf{m}}{}^{\alpha}{}_{\beta'}\psi_{\mathbf{n}]a\alpha} - \psi_{[\mathbf{m}}{}^b{}_{\beta'}A_{\mathbf{n}]ab})$$

$$\delta\tilde{\psi}_{\mathbf{m}}{}^a{}_{\alpha'} = \partial_{\mathbf{m}}\rho^a{}_{\alpha'}$$

$$\delta\widetilde{R}_{\mathbf{mn}a\alpha'} = \partial_{[\mathbf{m}}\kappa_{\mathbf{n}]a\alpha'},\quad \delta\tilde{\psi}_{\mathbf{m}}{}^a{}_{\alpha'} = \eta^{ab}\kappa_{\mathbf{m}b\alpha'}$$

While the non-self-dual ungauged N=8 theory had an $E_{7(+7)}$ global symmetry for the scalars, the self-dual ungauged N=8 theory has only a contraction of that. Besides the manifest SL(8) invariance, there is the usual invariance of free scalar theories of translation of the scalars by constants, which is here accompanied by a corresponding transformation of $G$:

$$\delta\phi_{abcd} = \sigma_{abcd},\quad \delta G_{abcdef\alpha'\beta'} = -\tfrac{1}{48}\sigma_{[abcd}F_{ef]\alpha'\beta'}$$

The proof of equivalence to the light-cone gauge action involves choosing an appropriate gauge and eliminating all auxiliary fields. (Again, the equivalence to string results follows from choosing a somewhat different gauge and treating a different set of fields as auxiliary, in the same way as for the superspace covariant derivatives. In the case of the heterotic string, a gauge is chosen where the antisymmetric part of the vierbein is identified with a gauge field resulting from dualizing one of the scalars, allowing a solution of the constraints for those fields which does not resemble that for the other gauges.) In some cases, eliminating auxiliary fields involves solving constraints by writing remaining fields as (space) derivatives of new fields; this was already done above in the derivation of the light-cone equations of motion from the commutators of covariant derivatives. One can then obtain the same action as that resulting from expanding the light-cone action over $\theta$. The complete proof of equivalence will require proof of local supersymmetry invariance of the action, so that the light-cone gauge choices will be justified. This equivalence is the simplest way to demonstrate that the light-cone action is Lorentz invariant (although the light-cone equations of motion have already been shown to be covariant).

For comparison, we give the covariant component action for self-dual N=4 super Yang-Mills (open string) [5] and the global supersymmetry transformations (which we have completely verified): The lagrangian is

$$L = \tfrac{1}{2}G^{\alpha'\beta'}F_{\alpha'\beta'} + \chi^{a\alpha'}\nabla^{\alpha}{}_{\alpha'}\chi_{a\alpha} + \epsilon^{abcd}(\tfrac{1}{8}\phi_{ab}\Box\phi_{cd} + \tfrac{1}{4}\phi_{ab}\chi_c{}^{\alpha}\chi_{d\alpha})$$

The supersymmetry transformations are

$$\delta A_{\alpha\alpha'} = \epsilon^a{}_{\alpha'}\chi_{a\alpha}$$
$$\delta \chi_{a\alpha} = \epsilon^{b\alpha'}\nabla_{\alpha\alpha'}\phi_{ba} \qquad\qquad -\epsilon_a{}^{\beta}F_{\alpha\beta}$$
$$\delta \phi_{ab} = -\epsilon_{abcd}\epsilon^{c\alpha'}\chi^d{}_{\alpha'} \qquad\qquad +\epsilon_{[a}{}^{\alpha}\chi_{b]\alpha}$$
$$\delta \chi^a{}_{\alpha'} = \tfrac{1}{2}\epsilon^{a\beta'}G_{\alpha'\beta'} + \tfrac{1}{2}\epsilon^e{}_{\alpha'}\epsilon^{abcd}[\phi_{bc},\phi_{de}] \quad +\tfrac{1}{2}\epsilon^{abcd}\epsilon_b{}^{\alpha}\nabla_{\alpha\alpha'}\phi_{cd}$$
$$\delta G_{\alpha'\beta'} = -\epsilon^a{}_{(\alpha'}\chi^b{}_{\beta')}\phi_{ab} \qquad\qquad +\epsilon_a{}^{\alpha}\nabla_{\alpha(\alpha'}\chi^a{}_{\beta')}$$